\documentclass[useAMS,usenatbib]{mn2e}

\newcommand{\mr}[1]{\mathrm{#1}}
\usepackage{graphicx}
\usepackage{txfonts}
\def\Mo{{\rm M}_\odot}
\def\Ro{{\rm R}_\odot}

\def\sfrac#1#2{{\textstyle\frac{#1}{#2}}}

\def\ud{\mathrm{d}}

\begin{document}
\title[Mass transfer in eccentric binaries]{Mass transfer in eccentric binaries:
the new Oil-on-Water SPH technique}
\author[R.~P. Church et al.]
{R.~P. Church$^{1,2}$ \thanks{E-mail: {\tt ross.church@sci.monash.edu.au}}, J. Dischler$^{1}$, M.~B. Davies$^{1}$, C.~A. Tout$^{2}$, T. Adams$^{3}$ and M.~E. Beer$^{3}$ \\
$^{1}$Lund Observatory, Box 43, SE-221 00, Lund, Sweden\\
$^{2}$Monash University, Mathematics Department, Clayton, Victoria 3168, Australia\\
$^{2}$Institute of Astronomy, The Observatories, University of Cambridge, Madingley Road, Cambridge. CB3 0HA\\
$^{3}$Department of Physics and Astronomy, University of Leicester, Leicester. LE1 7RH}

\date{Received;Accepted}

\pagerange{\pageref{firstpage}--\pageref{lastpage}} \pubyear{2008}

\maketitle

\label{firstpage}

\begin{abstract}
To measure the onset of mass transfer in eccentric binaries we have developed a
two-phase SPH technique.   Mass transfer is important in the evolution of
close binaries, and a key issue is to determine the separation at which mass 
transfer begins. The circular case is well understood and can be treated through
the use of the Roche formalism.  To treat the eccentric case we use a
newly-developed two phase system.  The body of the donor star is made up from
high-mass \emph{water} particles, whilst the atmosphere is modelled with
low-mass \emph{oil} particles.  Both sets of particles take part fully in SPH
interactions.  To test the technique we model circular mass-transfer binaries
containing a $0.6\,\Mo$ donor star and a $1\,\Mo$ white dwarf; such binaries are
thought to form cataclysmic variable (CV) systems.  We find that we can
reproduce a reasonable CV mass-transfer rate, and that our extended atmosphere
gives a separation that is too large by aproximately 16\%, although its pressure
scale height is considerably exaggerated.  We use the technique to measure the
semi-major axis required for the onset of mass transfer in binaries with a mass
ratio of $q=0.6$ and a range of eccentricities.  Comparing to the value obtained
by considering the instantaneous Roche lobe at pericentre we find that the
radius of the star required for mass transfer to begin decreases systematically
with increasing eccentricity.  \end{abstract}


\section{Introduction}
Mass transfer in binary stars is a very important process that completely
changes their evolution. Several eccentric binaries that
are undergoing mass transfer exist.  One such system is the low-mass X-ray
binary Cir X-1 \citep{murdin80,tauris99,johnston99,clarkson04}. Its orbital
parameters were analysed by \citet{tauris99} who argue that the system is
probably a $2 \, \Mo$ star orbiting a neutron star with an eccentricity of about
0.9.  Higher mass systems are more frequently observed; one such example is the
Be/X-ray transient A0538-66, located in the Large Magellanic Cloud
\citep{kretschmar04}.  This consists of a B2~IIIe star orbiting a neutron star
with a period of 16.65\,d in an eccentric orbit with $e \simeq 0.7$. In most of
these systems the mass transfer between the stars is driven by winds rather than
through Roche lobe overflow. 

Systems such as Cir X-1 are thought to be created when the heavier star explodes
in a supernova.  Another possible way to create a hard eccentric binary is
through a close encounter between a circular binary and a single star or binary.
The rate of such events is only significant when the stellar density is high, so
they occur primarily in environments such as the cores of globular clusters
\citep{davies95}. For a density of about $10^5\,\mr{stars/pc}^3$ the average
time between close encounters for a system is about 1\,Gyr. Hence most of the
binaries in a dense globular cluster core undergo at least one encounter.

The theory of mass transfer in the circular case, where the flow is continuous
and steady, has been studied for a long time and is well understood
\citep{paczynski71,renvoize02}.  For two stars in a circular orbit there exists
a corotating frame in which they are stationary and if the stars are taken to be
point masses the potential, known as the Roche potential, is well defined.  The
first connected equipotential surface that surrounds both stars is known as the
Roche lobe and provides a boundary for the potential well in which the star
sits.  If we make the assumption that the envelope of the star deforms to the
equipotential surface whilst conserving its volume then we can define the Roche
lobe radius such that a sphere with that radius has the same volume as the Roche
lobe \citep{eggleton83}.  This can then be used in conjunction with a
spherically symmetric model of the stellar structure, such as that produced by a
stellar evolution code, to predict the point of onset of mass transfer.  The
eccentric case is much more difficult because the assumptions made in order to
derive the Roche potential are no longer valid.  The relative velocities and
distances of stars in an elliptic orbit vary so there is no corotating frame.  

A commonly used method to assess whether mass transfer occurs in an eccentric
binary is to make the assumption that the stars corotate at periastron and apply
the circular theory there.  However it is not clear that it is possible to
define an instantaneous Roche lobe because, among other things, to do so assumes
that the timescale upon which the star adjusts to the changing force is
sufficiently short \citep{charles83,brown84}.

Another approach is to simulate such systems numerically. \citet{boyle86} made
simulations with test particles placed on the surface of the donor.
\citet{haynes80} simulated Cir~X-1 with a core and an extended atmosphere of
test particles. However they only looked at a single orbit. Full numerical
simulations of complete systems are rare but \citet{regos05} presented some
low-resolution smoothed particle hydrodynamic (SPH) simulations where they
considered four different eccentricities for a single semi-major axis $a$ and
mass ratio $q$. 

In this paper we introduce a new technique (called Oil-on-Water) within the SPH
formalism.  We define two types of SPH particles, heavy \emph{water} particles
that make up the stellar interior and very light \emph{oil} particles that sit
on top of the star. We are thus able to resolve the mass transfer even though
the fraction of the stellar mass transferred per orbit is very small.  We apply
this technique to mass transfer in eccentric binaries and investigate a variety
of binary systems, varying the semi-major axis $a$ and eccentricity $e$.

In Section~2 we discuss eccentric mass-transferring binaries from a theoretical
point of view, as well as the observational evidence for their existence. In
Section~3 the oil-on-water method is described and some tests presented. The
results of the simulations are given in Section~4 and discussed in Section~5.

\section{The Oil-on-Water technique}
In order to investigate mass transfer in eccentric binaries a two-phase SPH
technique has been developed. This section describes the new approach and some
tests of it.

\subsection{Smoothed particle hydrodynamics}
SPH was developed by \citet{lucy77} and \citet{gingold77}. 
It has been widely used in various astrophysical 
applications. SPH is a particle-based Lagrangian scheme: the particles'
motions follow the fluid velocity. This differs from grid based methods
where the fluid flow between a grid of cells is measured. An
advantage of SPH is that effort is not spent solving the hydrodynamic equations
in regions of space that are devoid of matter; in the case of a binary star
system simulated in a grid-based code this is most of the computational volume.
Another feature of SPH is that one can follow the evolution of the particles.
For reviews of SPH see \citet{benz90} and \citet{monaghan92}. Here only a few
key points will be mentioned in order to show how our implementation differs
from the standard technique. Our code is based upon the code of \citet{benz90}.

At the heart of SPH lies the kernel $W(\bmath{r},h)$ and the smoothing length
$h$. These set the size and shape of a particle's sphere of influence; how much
a particle affects other particles that lie at position $\bmath{r}$.  For
example the density at $\bmath{r}$ is given by
\begin{equation}
\label{eq_density}
\rho(\bmath{r}) = \sum_{{j}=1}^N  m_{j} W(|\bmath{r}-\bmath{r}_{j}|,h).
\end{equation}
One simply takes the sum of all the particles' masses weighted by
$W(|\bmath{r}-\bmath{r}_{j}|,h)$. If particle $j$ is far from $\bmath{r}$
its contribution to the sum is negligible (or even zero), whilst if it is close
it has a large contribution.

SPH is a Lagrangian averaging scheme so it is easy to take derivatives and 
hence simple to calculate the force on a particle $i$. For example the
derivative of the density is given by
\begin{equation}
\langle \nabla \rho(\bmath{r}) \rangle = \sum_{j=1}^N  \frac{m_j}{\rho_j} \nabla W(|\bmath{r}-\bmath{r}_j|,h).
\end{equation}
Use of a similar expression for the derivative of the pressure $P$ and
symetrisation leads to the acceleration of a particle~$i$,
\begin{equation}
\frac{\mathrm{d}\bmath{v}_{i}}{\mathrm{d}t} = -\sum_{{j}=1}^N  m_{j} \left(
\frac{P_{i}}{\rho_{i}^2}+\frac{P_{j}}{\rho_{j}^2}+ \Pi_{{ij}} \right) \nabla W(\bmath{r},h) - \nabla \Phi,
\label{Eq_forcebal}
\end{equation}
where $P_i$ is the pressure at particle $i$; the first term is the symmetrized
expression for the pressure gradient.  The artificial viscosity $\Pi_{ij}$
is introduced in SPH to improve the treatment of shocks. We use the standard
formulation for this by \citet{monaghan83}.  We utilise a simple polytropic
equation of state with $\gamma=5/3$.  Finally, $\nabla \Phi$ gives the
gravitational force owing to all the other particles. For a star in stable
hydrostatic equilibrium the gravitational force on a particle is balanced by the
pressure gradient.

Gravity is not a local force so all particles have to be taken into account
and hence a brute-force calculation scales as $N^2$ in the particle number.
This would mean that the method would lose much of its usefulness because a
large amount of computational effort would be used for the gravitational force
summation. We use instead the hierarchical tree method of \citet{benz90b} which
employs a binary tree and scales as $N \log N$.   

To deal better with the large range in densities that arise in simulations of
these kind we employ a variable smoothing length~\citep{benz90}.  The derivative
of the smoothing length is calculated as
\begin{equation}
\frac{\ud h}{\ud t} = \sfrac{1}{3}h\nabla\cdot\mathbf{v}.
\end{equation}
We modify the derivative thus calculated to keep the number of neighbours that
each particle has between 80 and 120, subject to a maximum $h$ of $0.1\,\Ro$.

\subsection{The two phase technique: oil-on-water}
A typical mass-transfer rate for a low-mass close binary is $10^{-8}$ to $10^{-9}
\, \Mo\,{\rm yr}^{-1}$ \citep{patterson84}.  This implies that, in a typical
orbit of a few days to a fraction of a day, about $10^{-10}$ to $10^{-12} \,
\Mo$ is transferred. To resolve this mass flow using equal mass particles
we would require at least $10^{12}$ of them and this is beyond what can be
accomplished today. Within SPH it is possible to use a range of particle masses
but if the mass range is large numerical problems arise. As an example we can
consider the density calculation of Equation ~\ref{eq_density}.
If a particle of unit mass comes just within the range of a light particle with
a mass of $10^{-4}$ the heavy particle dominates the density
completely, even when the kernel at that point is very small.  We
circumvent this problem by introducing a two-phase scheme. The interior of the
star is made up of heavy \emph{water} particles while the atmosphere contains
very light \emph{oil} particles. It is these particles that take part in mass
transfer. We also introduce an artificial force to keep the two different types
of particles apart.  This is calculated from the number gradient rather than
from $\nabla P$.

  \begin{figure}                                                  
  \centering                                                       
  \includegraphics[width=84mm]{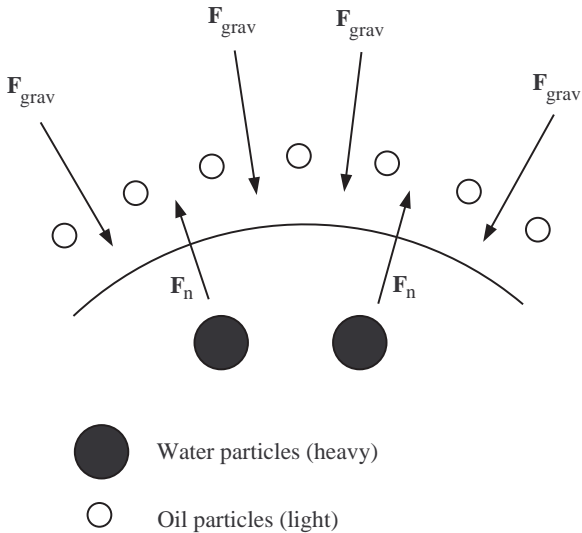}                          
     \caption{A cartoon figure of the force balance between the number gradient
     force $\bmath{F}_\mathrm{n}$ and the gravitational force
     $\bmath{F}_{\mathrm{grav}}$. The number gradient of the heavy water
     particles introduces a force perpendicular to the surface of the star. In
     order for this force to be smooth a sufficient number of water particles is
     needed. This is accomplished, in part, by a remapping of the water
     particles, see Section~\ref{sec_remap}}
\label{fig_OilWater}                                         
  \end{figure}       

In our simulations we use 15,390 water particles to simulate a $0.6 \, \Mo$
star so the average mass of a water particle is $3.92 \times 10^{-5} \,
\Mo$. The mass of each oil particle is $10^{-14} \, \Mo$ so the
average water particle has the same mass as $4 \times 10^{9}$ oil particles. We
use 32,691 oil particles in our runs, so the total mass in oil particles is negligible.

Figure~\ref{fig_OilWater} illustrates how the oil particles are balanced on top
of the water star. We have introduced an artificial force, $F_{\rm n}$, to
separate the oil layer from the water particles. This is based on the number
density of the water particles and therefore is perpendicular to the edge of the
water surface. This prevents the oil particles from penetrating into the stellar
interior and balances their gravitational attraction to the star.  It is defined
according to
\begin{equation}
\bmath{F}_\mathrm{n} = c_\mathrm{k} \nabla {n}_\mathrm{w}  , \qquad \nabla {n}_i = \sum_j \nabla_i W(|\bmath{r}-\bmath{r}_j|,h_\mathrm{ow}) 
\label{eq_fn}
\end{equation}
where the sum is over the water particles $j$.  To control the force we
introduce two parameters, $\alpha$ and $\beta$.  They are defined as
\begin{equation}
 h_\mathrm{ow} = \alpha h_\mathrm{w}
 \label{eq_alpha}
\end{equation}
\begin{equation}
c_\mathrm{k} = \beta \frac{G\Mo m_i}{\Ro^2}
\label{eq_beta}
\end{equation}
To calculate $\nabla {n}_i$ the smoothing length of the water,
$h_{\rm w}$ is used rather than the usual average between the particles. A small value
of $\alpha$ means that the oil and water particles must come closer to one
another before they begin to feel the repulsive force and so the oil layer lies
closer to the star. On the other hand the force must change smoothly otherwise a
typical oil particle will come towards the water particles with a large
velocity, bounce on the hard force barrier and get a large velocity kick
away from the star. A particle's time step decreases when the rate of
change of the force is large and this slows down the code. The strength of
the force is controlled by $\beta$ and the same arguments apply. Optimal values
for the parameters have been deduced from a large number of empirical tests. For
the production runs presented in this paper we use
\begin{equation}
\alpha = \sfrac{1}{2}
\end{equation}
and
\begin{equation}
\beta = 40.
\end{equation}

The force balance on the particles is then
\begin{equation}
\frac{\mathrm{d}{\mathbf v}_i}{\mathrm{d}t} = -\sum_{j=1}^N  m_j \left(
\frac{P_i}{\rho_i^2}+\frac{P_{j}}{\rho_{j}^2}+ \Pi \right) \nabla W(\mathbf{r},h)-
\frac{\mathbf{F}_{\rm n}}{m_{i}} - \nabla \Phi,
\end{equation}
where the number gradient force $F_{\rm n}$ is only applied for interactions
between oil and a water particles: see Table~\ref{tab:forcelaw} for further
clarification. The $F_{\rm n}$ force is similar to the normal pressure force
(see Equation~\ref{Eq_forcebal}) except that the acceleration it causes has no
dependence on the mass of the particles.  For interactions between two oil
particles or two water particles the ordinary equations are used. This is very
important because the oil particles are not test particles. They obey the full
SPH formalism so we can, for example, study the formation of an accretion disc
around the accreting star.

  \begin{figure}                                                  
  \centering                                                       
  \includegraphics[width=84mm]{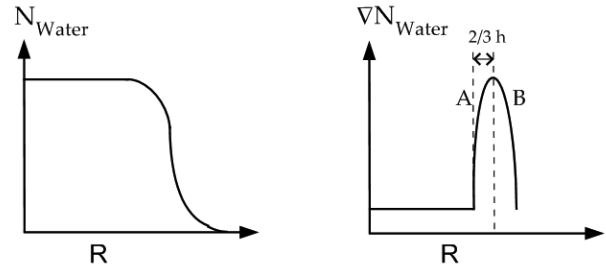}                          
     \caption{The left graph shows a schematic depiction of the number density
     of water particles computed using the SPH method.  The derivative of this
     is shown to the right.  If an oil particle comes within $\sfrac{2}{3}h$ of the
     water edge the repulsive force from the water particle density gradient
     declines and it can fall in towards the
     centre of the star. }             
        \label{fig_instab}                                          
  \end{figure}       

\begin{table}
\begin{tabular}{lll}
                          & Force on        & Force on \\
                          & water particle  & oil particle \\
                          \hline
Force from water particle & Standard        & Additional force $\mathbf{F}_{\rm n}$\\
Force from oil particle   & Standard        & Standard \\

\end{tabular}
\caption{The different force laws applied when calculating the force on a
particle due to one of the others.  Standard denotes the normal SPH force calculation.}
\label{tab:forcelaw}
\end{table}

\subsection{The oil-water kernel}
At the boundary between the oil and water particles the number density of the
water particles drops quickly (see Figure~\ref{fig_instab}, left panel).  This
gives rise to a problem with the number gradient force. Consider two oil
particles located at positions A and B in the right-hand panel. If the particle
at position B is perturbed slightly inwards the number density force 
increases and the particle is repulsed. However, if the particle at A moves
inwards the number gradient decreases and the infall continues. Therefore once a
particle is within $\sfrac{2}{3}h$ of the stellar surface it falls into the star.
This can be prevented by keeping the oil particles sufficiently far away from
the water edge. However this creates a separation between the oil particles and
the star. 

Another approach to this problem is to change the kernel for the oil-water
interaction when calculating $\mathbf{F}_{\rm n}$. The kernel used to calculate
the number gradient force does not have to be the same as that used for the rest
of the calculations. Since this interaction is very different the demands on
this kernel are different.  The most commonly used kernel is
\begin{equation}
W_{\rm ML}(r,h) =\frac{1}{\pi h^3} 
 \left\{
  \begin{array}{lccc}
    1-\frac{3}{2}v^2 + \frac{3}{4}v^3  &&&   0 \leq v \leq 1  \\
    \frac{1}{4} (2-v)^3                &&&   1 \leq v \leq 2  \\
    0                                  &&&  \mathrm{otherwise}
  \end{array}  
 \right.,
 \label{eq_mlkern}
\end{equation}
where $v = r/h$
\citep{monaghan85}.
Use of this kernel causes particles to clump together at a separation of $r =
\sfrac{2}{3}h$.  This is because the derivative of the kernel $\nabla W(r,h)$ has a
minimum at this point. In our case this would imply that the oil layer is 
$\sfrac{2}{3}\alpha h$ from the edge of the water star.  Because we aim to have the oil
layer as close to the star as possible we investigated three other kernels,
\begin{itemize}
\item a kernel without a {\it break point} (minimum in its derivative), 
\item the kernel of \citet{herant94} and
\item a kernel $W_p(r,h)$, constructed with a pair of polynomial functions, with
a break point at $v_{\rm b}=0.4$.
\end{itemize}

The kernel without a break point proved unsatisfactory because as particles come
close to one another the timestep falls sharply.  To prevent this a large
value of $c_{\rm k}$ must be used to prevent the oil particles from approaching the
star. Hence the oil particles are placed far from the water surface.  The other
two kernels were both well-behaved but $W_p(r,h)$ seemed somewhat more robust
and hence was used for our production runs presented in
Section~\ref{sec_results}.  It is given by

\setlength\arraycolsep{2pt}
\begin{equation}
W_p(r,h) =\frac{1}{\pi h^3} 
 \left\{
  \begin{array}{lccc}
    \frac{231}{142}-\frac{819}{142}v^2+\frac{1869}{284}v^3-\frac{315}{142}v^4  &&&   0 \leq v \leq 1  \\
    \frac{63}{284} (2-v)^3                                                     &&&   1 \leq v \leq 2  \\
    0                                                                          &&&  \mathrm{otherwise}
  \end{array}  
 \right..
 \label{pol04}
\end{equation}

To obtain this kernel, we first choose it to be the same as $W_{\rm ML}$ for
$r>h$, subject to a normalisation factor.  We construct a new fourth-order
polynomial for $r<h$, subject to the constraints that
\begin{itemize}
\item the kernel and its first and second derivatives are continuous at $v=1$,
\item the first derivative of the kernel is zero at $v=0$, 
\item the first derivative of the kernel contains only one extremum, the break
point, which is a minimum, and
\item the break point is located at $v_{\rm break}<\sfrac{2}{3}$ (we choose
$v_{\rm break}=0.4$).
\end{itemize}
These conditions specify the kernel $W_p$ up to the standard normalisation
factor.

\subsection{Remapping of the water particles} \label{sec_remap}
It is advantageous to have as many water particles as possible in the outer
parts of the star and hence close to the interface of the oil and water
particles.  This can be accomplished by remapping the water particles to
increase the number density at the edge of the star. The star is initially built
up using a close-packed lattice with the water particles placed equidistantly.
Particle masses are assigned to obtain the correct density profile, following
the YREC-models of \citet{guenther92}. To increase the concentration of water
particles at the surface we remap their positions according to a power law of
index $\gamma = 0.9$.  Each particle $i$ is updated according to
\begin{equation}
r_{i, \mathrm{remap}} = c_{\mathrm{norm}} r_i^{\gamma}
\end{equation}
where $c_{\mathrm{norm}}$ is a normalisation factor to keep the stellar radius
constant. This increases the number density at the outer edge by a factor of
about 1.25  with respect to the centre.  The larger number of water particles at
the edge makes the surface smoother; this allows us to place the oil layer
closer to the surface.

  \begin{figure*}                                                  
  \centering                                                       
  \includegraphics[width=175mm]{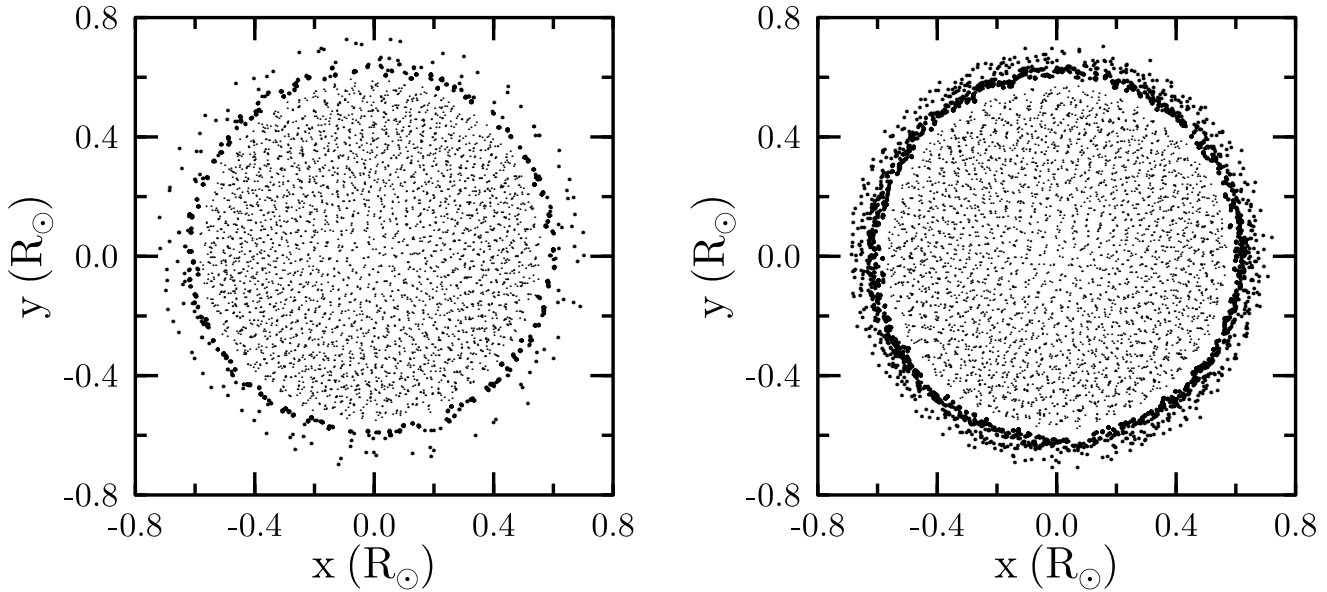}                          
     \caption{Two examples of a relaxed composite star. The particles plotted
     are those within $h$ of the $xy$ plane.  The small dots are water particles
     and the large dots oil particles.  In both models there are 15,390 water
     particles.  In the left-hand model there are 9,983 oil particles: in the
     right-hand 32,691.  In the right-hand figure a clear density structure can
     be seen in the oil particles.}                      
        \label{singlestar}                                          
  \end{figure*}                                             

\subsection{Production of the single-star models}
To build a star the water particles are first set up using the procedure
described above. The resulting model is then relaxed for approximately
100~$\tau_{\mr{relax}}$, where $\tau_{\mr{relax}}$ is the dynamical timescale
given by  
\begin{equation}
\tau_{\mr{relax}}=\sqrt{\frac{\Ro^3}{G\Mo}}
\end{equation}
Then the oil layers are added on top of the relaxed water star. We have tried
different numbers of oil particles: one and three layers of oil require 9,983
and 32,691 oil particles respectively. These are added according to a
close-packed lattice with a spacing equal to the internal smoothing length,
$h=0.025\,\Ro$.  The oil particles each have a mass of
$10^{-14}\,\Mo$.  The model is relaxed until the particles have settled
into a stable configuration; this typically takes about 50~$\tau_{\mr{relax}}$. 

The two stars with one and three layers of oil are shown in
Figure~\ref{singlestar}.  In the model that originally had a single layer the
oil particles are evenly spread around the star but there is no depth in the
oil layer. The other has visible structure.  The oil particles all have
the same mass so density is well-mapped by the number density of the oil
particles. There are more oil particles close to the water-oil boundary so
the atmosphere is densest here.  The simulations presented here all
utilise 32,691 oil particles as in the three-layer model. The remapping of the
water particles is also visible in the density contrast between the centre and
surface of the water particle star.

\section{Circular binaries}

In order to test the reliability of the oil-on-water model we used the code to
model circular binary systems.  Because the potential field is circular,
co-rotating binaries is well described by the Roche formalism there is a large
body of work that describes them and hence useful comparisons may be made.  Two
questions can be posed.  First, can we make a binary system that reproduces
observed properties correctly?  Second, how much does the unphysically large
size of the atmosphere in our models affect the results that we obtain?  We
answer these two questions by making models of a cataclysmic variable system in
the manner described below.

\subsection{The Roche potential}
\label{sec_rochelobe}
In a circular binary where one star is filling its Roche lobe, the mass flow is
continuous and steady, and a number of simplifications can be made.  By assuming
that the stars are centrally condensed for the purposes of calculating the
potential and that they are corotating with the orbit we obtain the Roche
potential \citep{pringle85}.  In a corotating frame centred on star 1, with the
line connecting the two stellar centres along the $x$ axis, the potential takes
the form
\begin{eqnarray}
\label{eq_rl_pot}
\lefteqn{ \Phi_R (x,y,z) = - \frac{G M_1}{\sqrt{x^2+y^2+z^2}} - \frac{G M_2}{\sqrt{(x-a)^2+y^2+z^2}}  } \nonumber \\ 
& & {}  - \frac{1}{2} \Omega_{\rm C}^2 \left[ (x-\mu a)^2 + y^2 \right],
\end{eqnarray}
where $a$ is the separation, $\Omega_{\rm C}$ the angular velocity and the
reduced mass $\mu =
M_2/(M_1+M_2)$. The last term in this expression is due to the centrifugal
force.  The stationary point between the two stars is the inner Lagrangian or L1
point, and the surface around each star that passes through this point is its
Roche lobe.  The radius of a spherical star with the same volume as
its Roche lobe is called the Roche-lobe radius $R_{\rm L}$ and is fitted
within a few percent by 
\begin{equation}
\frac{R_{\rm L}}{a} = \frac {0.49 q^{2/3}} {0.6 q^{2/3}+\ln(1+q^{1/3})},
\label{eq_eggleton}
\end{equation}
\citep{eggleton83}, where $q=M_2/M_1$ is the mass ratio. 
If one of the stars fills its Roche lobe mass flows
through L1 to the other star.  \citet{ritter88}
showed that the amount of mass transferred in a binary system depends on the
pressure scale height $H_{\rm p}$ of the donor star's atmosphere and the
difference between the donor star's radius $R$ and $R_{\rm L}$, according
to
\begin{equation}
\dot{M} = \dot{M}_0 e^{-(R_\mr{L}-R)/H_\mr{p}},
\label{eqn:ritter}
\end{equation}
where $\dot{M}_0$ is the mass-transfer rate of a binary that just fills its
Roche lobe.  For a low-mass main-sequence star $\dot{M}_0\simeq10^{-8}\,\Mo
\,{\rm yr}^{-1}$ and $H_{\rm p} \simeq 10^{-4}\,\Ro$.  

\subsection{Production of circular binary models}

The transition from the spherical potential produced by the self-gravity of a
single star and the Roche potential owing to an orbiting binary system disrupts
the structure of our relaxed star unless care is taken. For this reason it is
necessary to relax a binary system once the star has been placed in its orbit
before allowing mass transfer to proceed.  To do this we first add a point mass
to the star to represent a compact companion.  In all these runs we have used a
star with mass $0.6\,\Mo$ and a $1\,\Mo$ point mass.  Initially we place the
star and point mass in a frame rotating about the centre of mass with a damping
force in place to allow the star to adjust to the gravitational field of the
point mass.  The star is taken to be co-rotating: it is stationary in the
rotating frame other than for oscillations during relaxation.  Once the star has
relaxed the model and point mass are transformed back into an inertial frame.
The star and the point mass are both given the correct velocities to place them
on the desired orbit.  Relaxation takes approximately $10\,\tau_{\rm relax}$.  

For an eccentric mass-transfer binary the relaxation takes place in a circular
orbit at the apocentre separation where no mass transfer takes place.  In
contrast for a circular orbit the separation of the stellar centre and point
mass is constant and hence there is no point in the orbit at which the
relaxation can take place without mass transfer occurring.  To avoid this problem
we relax the star in a circular orbit that is slightly wider than the widest
orbit in which appreciable mass transfer takes place.  After relaxation the
separation of the orbit is reduced by 0.5\% each dynamical timescale, changing
the positions and velocities of the particles.  This provides a series of
circular models at different separations that can be evolved further to model a
series of binaries.

At low mass-transfer rates we encounter numerical problems.  Because the rate of
flow of particles into the accretor's Roche lobe is small there are only a few
particles present, which causes the pressure forces to vary very rapidly.  
These particles gain large velocities and are ejected from the Roche lobe.  To
counter this problem we turn off sph forces for oil particles with fewer than 50
neighbours: that is, such particles move balistically.

\subsection{Results}
\begin{figure}
   \centering
   \includegraphics[width=\columnwidth]{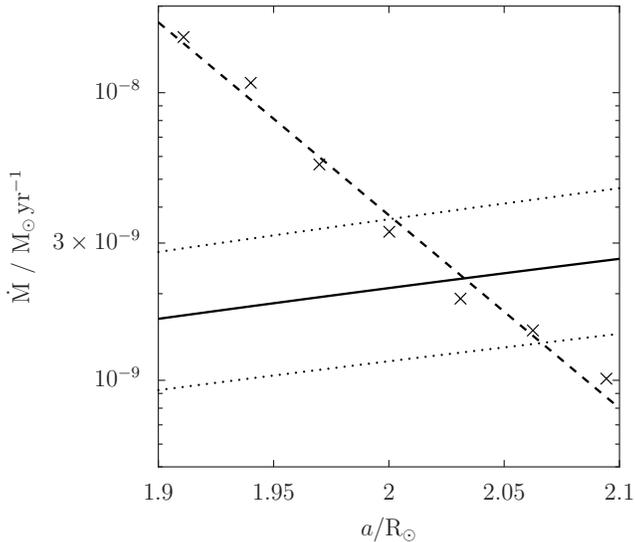}
   \caption{The mass-transfer rate in our circular binary models.  The crosses
   show the rate at which mass is accreted in the circular systems and the 
   dashed line is a straight-line fit to these points.   The solid line is the
   observed accretion rate for CVs from Equation~17 of \citet{patterson84}, and
   the dotted lines its 1-sigma error interval.}
   \label{fig:circaccretion}
\end{figure}

All the circular binaries consisted of a $0.6\,\Mo$ star, modelled using the
two-phase oil-on-water technique and a $1\,\Mo$ point mass.  Such binaries are
believed to be relatively high-mass cataclysmic variable (CV) systems. In these
hydrogen-rich material accretes on to a white dwarf and burns explosively
at the surface.  Ongoing mass transfer is thought to be driven by angular
momentum loss owing to either a magnetic wind or gravitational radiation.
Observations of many such systems have been made and their accretion rates can
be measured from their luminosities.  We ran models at a range of separations,
which yielded a corresponding range of mass-transfer rates.  We measure the rate
by fitting the number of particles in the Roche lobe of the accretor as a linear
function of time, using a standard least-squares fitting algorithm.  The
accretion rate for a given binary was roughly constant until a substantial
fraction of the envelope had been transferred.  The accretion rates we obtain
are shown in Figure~\ref{fig:circaccretion}, along with an observational
measurement of the CV accretion rate from~\citet{patterson84}.

Mass transfer from a CV is driven by a loss of angular momentum from the system
and the mass-transfer rate is seen to be a function of the binary period.  At
the points where the observed accretion rates and the accretion rates obtained
from our code are equal we can take the separation to be that predicted for the
CV we are modelling.  We find this equilibrium separation to be
$a=2.03^{+0.04}_{-0.03}\,\Ro$.
Applying Equation~\ref{eq_eggleton} with $q=0.6$ gives a Roche lobe radius
$R_{\rm L} = 0.68^{+0.015}_{-0.01}\,\Ro$.  The mean volume of the star, as
defined by the volume occupied by the water particles, was estimated by finding
its extreme points along the three natural axes of the system and calculating
the volume of an ellipsoid with those axis lengths.  The radius of the sphere
with the same volume, in analogy to the definition of the Roche lobe radius, is
$0.57\,\Ro$.  If we assume that the real edge of the star coincides with the
outer surface of the water particles this gives an error in the effective radius
of the stellar model owing to the extended atmosphere of 15 to 18\%.

The mass-transfer rate in our models varies in an approximately exponential fashion with
separation.  Hence the model shows the qualitative behaviour predicted by
\citet{ritter88}, which suggests that the atmosphere is sufficiently resolved.  By
rearranging Equation~\ref{eqn:ritter} and substituting for the Roche lobe radius
from Equation~\ref{eq_eggleton} with $q=0.6$ we obtain
\begin{equation}
\log\left(\frac{\dot{M}}{\dot{M}_0}\right) = \frac{R}{H_{\rm p}}-0.3356\frac{a}{H_{\rm p}}.
\label{eqn:Ritter2}
\end{equation}
Equating the final term to the variation in accretion rate from
Figure~\ref{fig:circaccretion} gives an effective pressure scale height of
$0.022\,\Ro$, unsurprisingly very similar to the smoothing length $h$.  The
pressure scale height of a low-mass main-sequence star is about $10^{-4} \,
\Ro$, roughly two orders of magnitude smaller than in our models.  This
means that the onset of mass transfer in our models will be much gentler than
that seen in real systems.

  \begin{figure*}                                                  
  \centering                                                       
  \includegraphics[width=175mm]{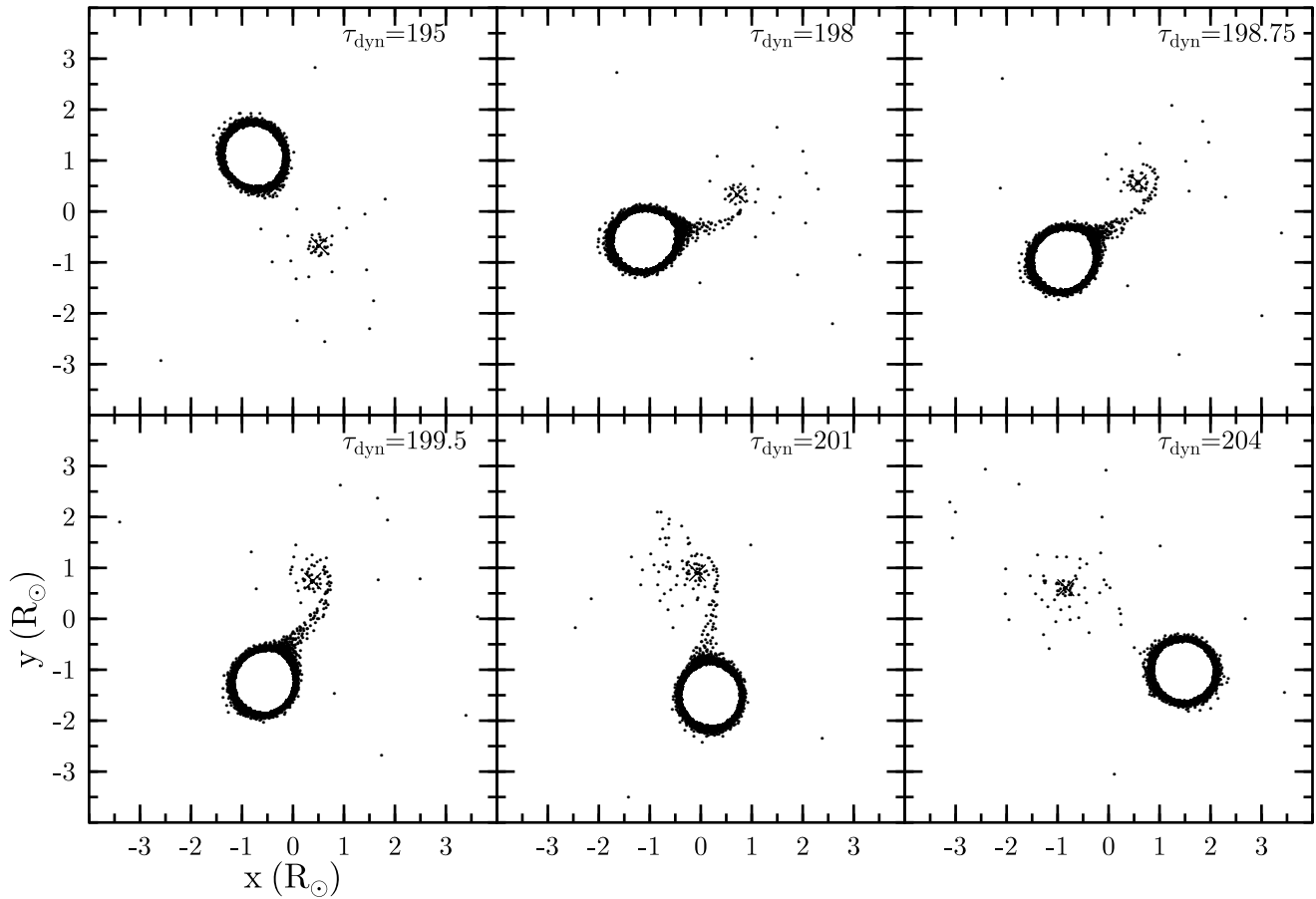}                          
     \caption{An example of mass transfer in an eccentric binary with $a=2.5
     \, \Ro$ and $e=0.2$. Here for clarity only oil particles within $h$ of the
     $xy$-plane are plotted. The position of the point-mass companion
     is marked with a cross. The mass of the star is $0.6 \, \Mo$ and that
     of the point mass $M=1.0 \, \Mo$, as in all the simulations in this
     paper.  In this binary mass is transferred only near periastron, which occurs
     when the star is to the left of the point-mass along the $x$-axis. This image
     sequence starts just before periastron passage and, as the star approaches 
     periastron the mass transfer begins; however there is a noticeable lag.  The
     mass transfer actually begins just after the closest approach and continues
     for almost 1/4 of the orbit.}                      
        \label{fig_movie}                                          
  \end{figure*}                                             

\begin{table}                                                                                    
  \centering                                                                                     
    \caption{A table of the runs with different semi-major axes $a$ and
    eccentricities $e$. In all cases the mass ratio $q=0.6$. Column~4 indicates
    whether no mass transfer (N), mass transfer (Y) or massive mass transfer (M)
    was observed. For the mass-transfer case the number of particles escaping
    the donor per orbit is given in Column~5. }
\label{tab_runs}                                                                                     
\begin{tabular}{|l|l|l|l|l}                                                                        
  \hline                                                                                         
  $a/\Ro$   &  $e$   &  $r_{\mr{peri}}/\Ro$  &  Mass transfer  & Oil particles lost  \\
                       &        &                   &                 & from donor per orbit  \\
  \hline                                                                           
  2.5		       & 0.0    & 2.500               & N               & -    \\
  2.5                  & 0.15   & 2.125              & Y               & 30   \\
  2.5                  & 0.20   & 2.000             & Y               & 200  \\
  2.5                  & 0.25   & 1.875              & Y               & 500  \\
  &&&\\                                            
  3.0                  & 0.0    & 3.000               & N               & -    \\
  3.0                  & 0.15   & 2.550              & N               & -    \\
  3.0                  & 0.30   & 2.100              & Y               & 20   \\
  3.0                  & 0.35   & 1.950              & Y               & 200  \\
  3.0                  & 0.375  & 1.875             & M               & -    \\
  3.0                  & 0.40   & 1.800             & M               & -    \\
  &&&\\                                            
  4.0                  & 0.40   & 2.400              & N               & -    \\
  4.0                  & 0.45   & 2.200              & N               & -    \\
  4.0                  & 0.50   & 2.000              & Y               & 40   \\
  4.0                  & 0.525  & 1.900             & Y               & 250  \\
  4.0                  & 0.55   & 1.800              & M               & -    \\
  &&&\\                                            
  5.0                  & 0.50   & 2.500              & N               & -    \\
  5.0                  & 0.60   & 2.000              & Y               & 10   \\
  5.0                  & 0.625  & 1.875             & M               & -    \\
  5.0                  & 0.65   & 1.750              & M               & -    \\
  &&&\\                                            
  6.0                  & 0.65   & 2.100               & N               & -    \\
  6.0                  & 0.675  & 1.950             & Y               & 50   \\
  6.0                  & 0.6875 & 1.875            & M               & -    \\
  6.0                  & 0.70   & 1.800             & M               & -    \\
  &&&\\                                            
  7.0                  & 0.70   & 2.100              & N               & -    \\
  7.0                  & 0.75   & 1.750              & M               & -    \\
  &&&\\                                            
  8.0                  & 0.75   & 2.000              & Y               & 60   \\ 
   \hline                                                                                         
\end{tabular}                  
\end{table}                                                             

\section{Eccentric binaries}
\label{sec_results}

To simulate eccentric binaries we carried out a grid of runs with different $a$
and $e$.  In each case the donor was a main-sequence star of $M = 0.6 \,
\Mo$ and the accreting companion was a white dwarf with $M = 1 \,
\Mo$.  To relax the star the point mass was initially placed at a distance
$a_{\mathrm{apo}}$ from the star in a circular orbit as explained above.  When
the system had relaxed sufficiently we released it from co-rotation and 
introduced eccentricity in such a way that it was relaxed at apocentre: the
spin of the star was set to coincide with the angular velocity of the system
there.  Therefore the star initially co-rotates at apastron. The spin rate is
not found to change significantly during a complete run.
Figure~\ref{fig_movie} is an example of a system with $a=2.5 \, \Ro$ and
$e=0.2$ which undergoes mass transfer only at periastron, so the mass transfer
turns on and off in each orbit.  One such mass-transfer event is shown.

Table~\ref{tab_runs} enumerates parameters for the grid of runs completed,
stating whether mass transfer takes place and, if so, the number of
oil particles escaping the donor star per orbit. More eccentric systems have a
narrower range of $a$ in which stable mass transfer takes place. This is
expected because the periastron separation is given by $a_{\rm p}=a(1-e)$, so a
small change in $e$ leads to a large change in the periastron separation. 

  \begin{figure}                                                  
  \centering                                                       
  \includegraphics[width=84mm]{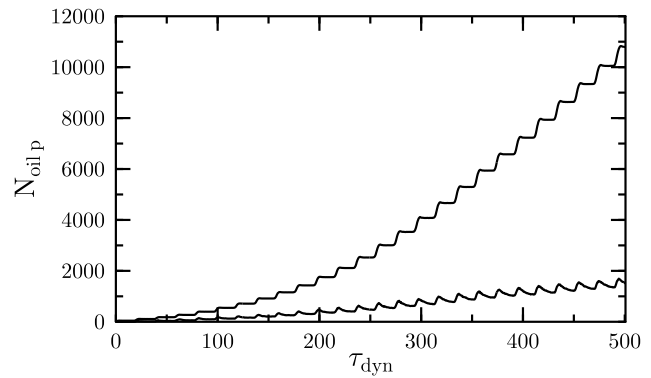}                          
     \caption{The cumulative number of oil particles released from the donor
     against time in the case of
     an eccentric system ($a=2.5\,\Ro$, $e=0.2$). After the first
     initialisation, where a damping force is active, there is a stable flow of
     particles at periastron and this gives raise to the stepped 
     function. The lower curve shows the number of particles captured by
     the companion. Most of the particles transferred are not captured but
     escape through L2. As discussed in section~\ref{sec_rochelobe} the
     potential at L2 is almost the same height as the inner Lagrangian point L1
     for $e=0.25$}.
        \label{fig_mt_a25e20}                                          
  \end{figure}                                             

  \begin{figure}                                                  
  \centering                                                       
  \includegraphics[width=84mm]{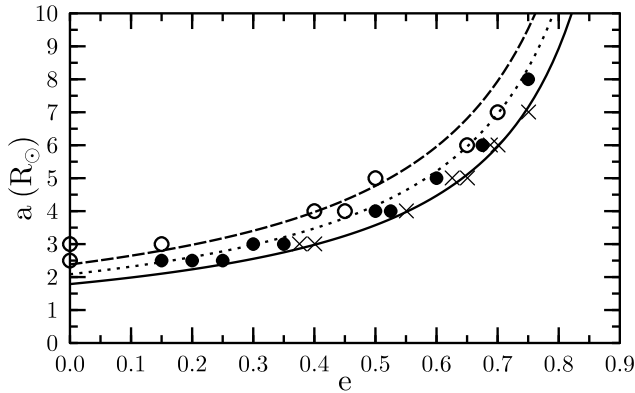}
  \caption{Results from runs at different $a$ and $e$. Open circles denote
  binaries where no mass transfer was seen, filled circles are runs with stable
  mass transfer, and crosses mark models where a massive overflow of particles
  was observed and water particles began to transfer. The curves show the
  separations derived for the onset of mass transfer under the assumption that
  the star exactly fills its Roche lobe at periastron.  The three curves are for
  stellar radii of 0.6, 0.7 and 0.8 $\Ro$, with $0.8\,\Ro$ being the
  topmost curve. }                      
        \label{fig_mt_mine}                                          
  \end{figure}                                             

An example of the mass transfer in one of our runs is given in
Figure~\ref{fig_mt_a25e20}. The periodic behaviour of the
mass transfer is apparent. On each orbit mass begins to flow from the donor at
pericentre. The number of oil particles lost from the donor at each periastron
passage is roughly constant after the first few orbits.  This indicates that the
mass transfer is stable. One periastron passage of this system is also pictured
in  Figure~\ref{fig_movie}. 

Our models are plotted in the ($a,e$)-space in Figure~\ref{fig_mt_mine}. Open
circles denote systems without mass transfer, filled circles stable mass
transfer and crosses denote that the there is a massive overflow of particles
with water particles being transferred.  The prediction of
Equation~\ref{eq_eggleton} for an instantaneous Roche lobe at pericentre is also
plotted.  Such an approach assumes that the timescale of the mass transfer is
sufficiently fast to be able to adjust to the instantaneous distance and
rotational velocity. The typical orbital passage shown in Figure~\ref{fig_movie}
shows that this is not the case.  The mass transfer does not begin until just
after periastron passage and continues for almost 1/4 of an orbit thereafter.
This is due to the time taken for the material to respond to the changing
potential as the star goes round the periastron passage and the time taken for
matter to flow from the donor to the companion. The instantaneous Roche lobe
construction also assumes corotation, which only can be true at one point in the
orbit.

  \begin{figure}                                                  
  \centering                                                       
  \includegraphics[width=84mm]{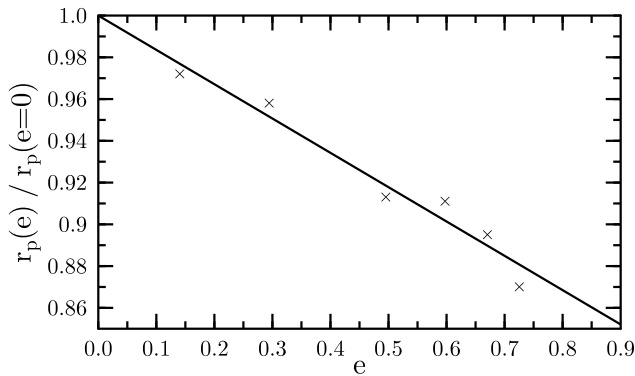}                          
     \caption{The ratio $r_{\mr{p}}(e) / r_{\mr{p}}(e=0)$ as a
     function of the eccentricity. The separation required for onset of mass
     transfer for different eccentricities is $r_{\mr{p}}(e)$ The line is a
     linear fit to the points, given by $f(e)=1-0.16e$. }                      
        \label{fig_eggletonkorr}                                          
  \end{figure}                                             

In order to quantify the effect of the eccentricity we express the periastron
distance required for onset of mass transfer, $r_{\mr{p}}(e)$, as
\begin{equation}
\label{eq_eggletokorr}
\label{eq_111}
r_{\mr{p}}(e) = r_{\mr{p}}(e=0) \times f(e),
\end{equation}
where $r_{\mr{p}}(e=0)$ is the separation required for the circular case and
$f(e)$ is a correction function to be determined.  If the instantaneous Roche
lobe formalism was applicable, i.e. the onset of mass transfer was determined by
the Roche lobe at periastron, we would obtain $f(e)=1$.  Based on our runs in
Table~\ref{tab_runs} we have  estimated the eccentricity for which the mass
transfer begins for a given semi-major axis $a$.  The results are plotted in
Figure~\ref{fig_eggletonkorr}.  We have extrapolated the data to $e=0$ to obtain
an estimate of the periastron distance at which mass transfer starts in the
circular case as $r_{\mr{p}}(e=0)=2.212\,\Ro$ .
The required distance for onset of mass transfer between the stars can be seen
to decrease with increasing eccentricity.  The line is a linear fit to the
points which yields $f(e)=1-0.16e$.  Equation~\ref{eq_111} expressed in
terms of the semi major axis $a$ gives
\begin{equation}
\label{eq_eggletokorr2}
\label{eq_112}
a_{\mr{mt}}(e) = a_{\mr{mt}}(e=0) \times \frac{1-0.16e}{1-e},
\end{equation}
where $a_{\mr{mt}}(e)$ is the required semi major axis for mass transfer and
$a_{\mr{mt}}(e=0)$ can be obtained from Equation~\ref{eq_eggleton}.  Note
however that in our simulations all runs are corotating at apastron and $q=0.6$,
so the formula given above is only tested for this case. 

\section{Conclusions and future prospects}
The simulation of mass transfer in eccentric binaries is non-trivial because the
mass transferred in each orbit is a tiny fraction of the total mass. In this
article we have presented a technique that treats this problem by introducing
a two-phase SPH formalism.  Very light particles make up the outer part of the
donor star, while the inner part is formed from heavier particles.  Thanks to
this oil-on-water model have we been able to simulate mass transfer in eccentric
binaries. The simulations presented here cover a large range of semi-major axes
and eccentricities at a mass ratio of 0.6.

We have found that the onset of mass transfer in our simulations does not follow
the prescription of the circular case. We have measured the eccentricity
required for mass transfer given a semi-major axis $a$.  Using this we show that
the minimum distance between the stars that leads to mass transfer decreases
linearly with eccentricity. 

The oil-on-water model could be used to simulate other interesting astrophysical
processes. One example is the onset of common envelope evolution. A close binary
system undergoes a common envelope (CE) phase if the secondary star is unable to
accrete all the mass transferred by the donor \citep{paczynski76,rasio96}. The
core of the evolved star and the secondary orbit one another in a cloud of gas,
which is eventually expelled. This transports angular momentum away from the
system and the period of the system decreases. The oil-on-water technique can
follow the angular momentum transferred at the onset of the common envelope
phase accurately, making it an useful tool for such an investigation.

\section*{Acknowledgements}
The authors would like to thank the referee, Phil Armitage, for his constructive
comments on the manuscript.  RPC would like to thank the Swedish Institute for a
Guest Scholarship.  MBD is a Royal Swedish Academy Research Fellow supported by
a grant from the Knut and Alice Wallenberg Foundation.  CAT thanks Churchill
College for a Fellowship.

\bibliographystyle{mn2e}
\bibliography{oow2}

\label{lastpage}
\end{document}